\documentclass[pra,10pt,twocolumn,superscriptaddress,floatfix,showpacs]{revtex4-1}

\usepackage[utf8]{inputenc}  
\usepackage[T1]{fontenc}     
\usepackage[british]{babel}  
\usepackage[sc,osf]{mathpazo}
\usepackage[scaled=0.86]{berasans}  
\usepackage[colorlinks=true, allcolors=blue, urlcolor=blue]{hyperref}  
\usepackage{graphicx} 
\usepackage[babel]{microtype}  
\usepackage{amsmath,amssymb,amsthm,bm,amsfonts,mathrsfs,bbm} 

\usepackage{xspace}  
\usepackage{pgfplots}
\usepackage{xcolor,colortbl}
\usepackage{array}
\usepackage{bigstrut}
\usepackage{slashbox}

\newcommand{\ket}[1]{| #1 \rangle}
\newcommand{\bra}[1]{\langle #1 |}

\renewcommand{\rho}{\varrho}

\newcommand{\processnext}[1]{%
	\ifx\listfinish#1\empty\else\listact{#1}\expandafter\processnext\fi}

{\begin{framed}\begin{small}}
		{\end{small}\end{framed}}

\DeclareGraphicsExtensions{.pdf,.png,.jpg}

\begin{document}
	
	\title{Geometry of the quantum set on no-signaling faces}

	\author{Ashutosh Rai}
	\affiliation{International Institute of Physics, Federal University of Rio Grande do Norte, 59070-405 Natal, Brazil}
	
	\author{Cristhiano Duarte}
	\affiliation{International Institute of Physics, Federal University of Rio Grande do Norte, 59070-405 Natal, Brazil}
	\affiliation{Schmid College of Science and Technology, Chapman University, One University Drive, Orange, California, 92866, USA}
	
	\author{Samuraí Brito}
	\affiliation{International Institute of Physics, Federal University of Rio Grande do Norte, 59070-405 Natal, Brazil}
	
	\author{Rafael Chaves}
	\affiliation{International Institute of Physics, Federal University of Rio Grande do Norte, 59070-405 Natal, Brazil}
	\affiliation{School of Science and Technology, Federal University of Rio Grande do Norte, 59078-970 Natal, Brazil}
	
	
	\begin{abstract}
		Since Bell's theorem we know that quantum mechanics is incompatible with local hidden-variable models, the phenomenon known as quantum nonlocality. However, despite steady progress over the years, precise characterization of the set of quantum correlations remained elusive. There are correlations compatible with the no-signaling principle and still beyond what can be achieved within quantum theory, which has motivated the search for physical principles and computational methods to decide the quantum or postquantum behavior of correlations. Here, we identify a feature of Bell correlations that we call quantum voids: faces of the no-signaling set where all nonlocal correlations are postquantum. Considering the simplest possible Bell scenario, we give a full characterization of quantum voids, also understanding its connections to known principles and its potential use as a dimension witness. 
		
	\end{abstract}
	
	
	\maketitle
	
	
	\section{Introduction}
	Bell's theorem \cite{bell} is considered among the most fundamental discoveries in the foundations of quantum physics. Overall, Bell-type theorems prove that correlations generated between two (or more) noncommunicating spacelike separated parties, and that violate certain bounds, cannot be described through local hidden-variable models. In turn, correlations violating such bounds are termed as Bell nonlocal or simply \emph{nonlocal correlations} (see Ref.~\cite{review} for a review). Remarkably, quantum mechanical correlations can supersede local hidden-variable bounds, a phenomenon called \emph{quantum nonlocality}, and experiments performed to date have confirmed such violations in agreement with the predictions of quantum mechanics (Refs.~\cite{exp1,exp2,exp3} account for recent loophole free experiments).

	It is well known that nonlocal quantum correlations respect the no-signaling 
	principle~\cite{pr,bar}, basically stating that spacelike separated parties cannot directly influence each other's measurement statistics. Strikingly, however, there are nonsignaling correlations beyond what can be achieved within quantum theory, i.e., correlations compatible with special relativity but of a postquantum nature~\cite{pr}. Since this realization, research in the foundations of quantum mechanics has not only been concerned with the classical-quantum separation---witnessed by a violation of Bell inequalities---but also with the quantum-postquantum separation at the boundary of a set of quantum correlations~\cite{pop}.
	
	From a more foundational perspective, a number of (quantum) theory-independent physical principles have been proposed with the goal of explaining the boundary of the set of quantum correlations. Namely, the principles proposed so far are as follows: nontrivial communication complexity~\cite{ntcc}, no advantage for nonlocal computation~\cite{lnc}, information causality and its generalization~\cite{ic,ic-review,generalized-ic}, macroscopic locality~\cite{ml}, local orthogonality~\cite{lo}, and many-box locality 
	(a refinement of macroscopic locality)~\cite{mbl}. All of these principles provide bounds on the set of quantum correlations while reproducing some of its boundary points. For instance, by applying most of these principles~\cite{ic,ic-review,generalized-ic,ml,lo,mbl}, the Tsirelson bound \cite{tsirelson}, i.e., the maximum value for the paradigmatic Clauser-Horne-Shimony-Holtz (CHSH) inequality~\cite{chsh}, is reproduced. On the more applied side, points on the quantum boundary are shown to have direct relevance in device-independent quantum information processing~\cite{diqip,bbqm}, a novel framework where the successful execution of quantum protocols can be guaranteed by measurement statistics alone, without any need for assumptions about the physical systems being measured or the nature of these measurements.
	
	Covering both foundational and applied perspectives, a crucial aspect to better 
	understand quantum correlations, their potential advantages over classical resources but also their limitations in the processing of information, relies on understanding their geometry~\cite{goh}. Many more works~\cite{uffink,pitowsky,allcock1,allcock2,lang,Bierhorst,sam,GuRuJee,cris} 
	have also revealed a number of interesting geometrical aspects of the set of quantum correlations. Perhaps the best available tool for studying the quantum-postquantum boundary is the Navascues-Pironio-Acin (NPA) hierarchy~\cite{npa1,npa2}, which gives a series of outer approximations converging to a set of \emph{quantum correlations} $Q$. Interestingly, in a more recent development~\cite{aq}, it was shown that any nonlocal correlation which belongs to the set of \emph{almost quantum correlations} $Q^{(1+ab)}$, the set determined by the $(1+ab)$ level of the NPA hierarchy~\cite{npa1,npa2}, satisfies all physical principles proposed so far, with two possible exceptions: (i) the information causality principle~\cite{ic,ic-review} and its generalization~\cite{generalized-ic}, and (ii) the recently proposed principle of many-box locality~\cite{mbl}. Thus, with our  present understanding and from the known inclusion $Q\subsetneq Q^{(1+ab)}$ \cite{aq}, we can say that any postquantum correlation inside the almost quantum set cannot be fully detected by known physical principles.
	
	Here, it is worth mentioning that apart from hierarchically approximating or reproducing the set of quantum correlations from theory-independent physical principles, some important results on the geometry of the quantum set have been obtained in the theory-dependent framework. These results are either derived from the very mathematical structure of the quantum mechanics itself~\cite{cirelson,landau,mas1,mas2,wolfe,ishizaka1,ishizaka2} or by studying the interrelation of nonlocal quantum correlations with the uncertainty principle~\cite{uncertainity}, the complementarity principle~\cite{complimantarity}, and  measurement compatibility~\cite{compl-uncer} (for a 
	recent review, see Ref.~\cite{rev2}). Summing up, it is fair to say that although progress 
	has been made over the years, we still have a limited and fragmented understanding of 
	the quantum set of correlations.
	
	In this paper, we investigate a very rich and yet almost unexplored region of the quantum set of correlations, those lying on the faces of the no-signaling set~\cite{face}. For our purpose it is enough to consider the simplest Bell scenario. Applying the known symmetries in the CHSH scenario~\cite{bar}, we focus on analyzing one of the eight symmetries: the region defined by the convex hull of the (canonical) Popescu-Rohrlich (PR) box \cite{pr} and 
	eight deterministic local boxes on the corresponding local face. This gives us 
	an eight-dimensional nonlocal simplex whose no-signaling faces are simplexes of 
	dimension seven or less \cite{simplex}. Within this picture we define the concept of a \emph{quantum void}, i.e., regions on the faces of the no-signaling set where all nonlocal points are postquantum. As we show, faces of dimension six or less can give rise to quantum voids. As it turns out, all the no-signaling faces of dimension four or less are quantum 
	voids. Moving on, we analyze what some of the known physical principles and the 
	NPA$(1+ab)$ outer approximation of the quantum set have to say about the quantum voids. We find out that while principles such as no violation of information causality and macroscopic locality can reproduce some of the lower-dimensional quantum voids, nevertheless, even the set of almost quantum correlations $Q^{(1+ab)}$ cannot reproduce the six-dimensional quantum voids, and thus 
	none of the known physical principles can (with our actual knowledge) reproduce such sets of postquantum correlations. On the applied side, through an example, we show that quantum voids can have potential applications as device-independent dimension witnesses~\cite{dw1,dw2}.

	In addition, we analyze some of the no-signaling faces which are not quantum voids. Among these, we study the quantum Hardy nonlocal points~\cite{hardy,jordan,kar,rabelo,siba} which live on some of the five-dimensional faces. Considering the known results~\cite{ic-hardy1,ic-hardy2,ml-hardy,lo-hardy} showing the limitations of information causality, macroscopic locality, and local orthogonality 
	principles in reproducing the maximally nonlocal quantum Hardy point 
	\cite{rabelo,siba}, we computed a lower bound on the maximal success probability of Hardy's argument over the set of almost quantum correlations and find that the lower bound is larger than the maximum quantum value. Our result shows that all the known physical principles cannot (with our actual knowledge) reproduce the maximum success probability of Hardy's nonlocality argument in quantum mechanics.

	\section{PRELIMINARIES} \label{prelim}
\begin{figure}[h]
		\includegraphics[scale=0.30]{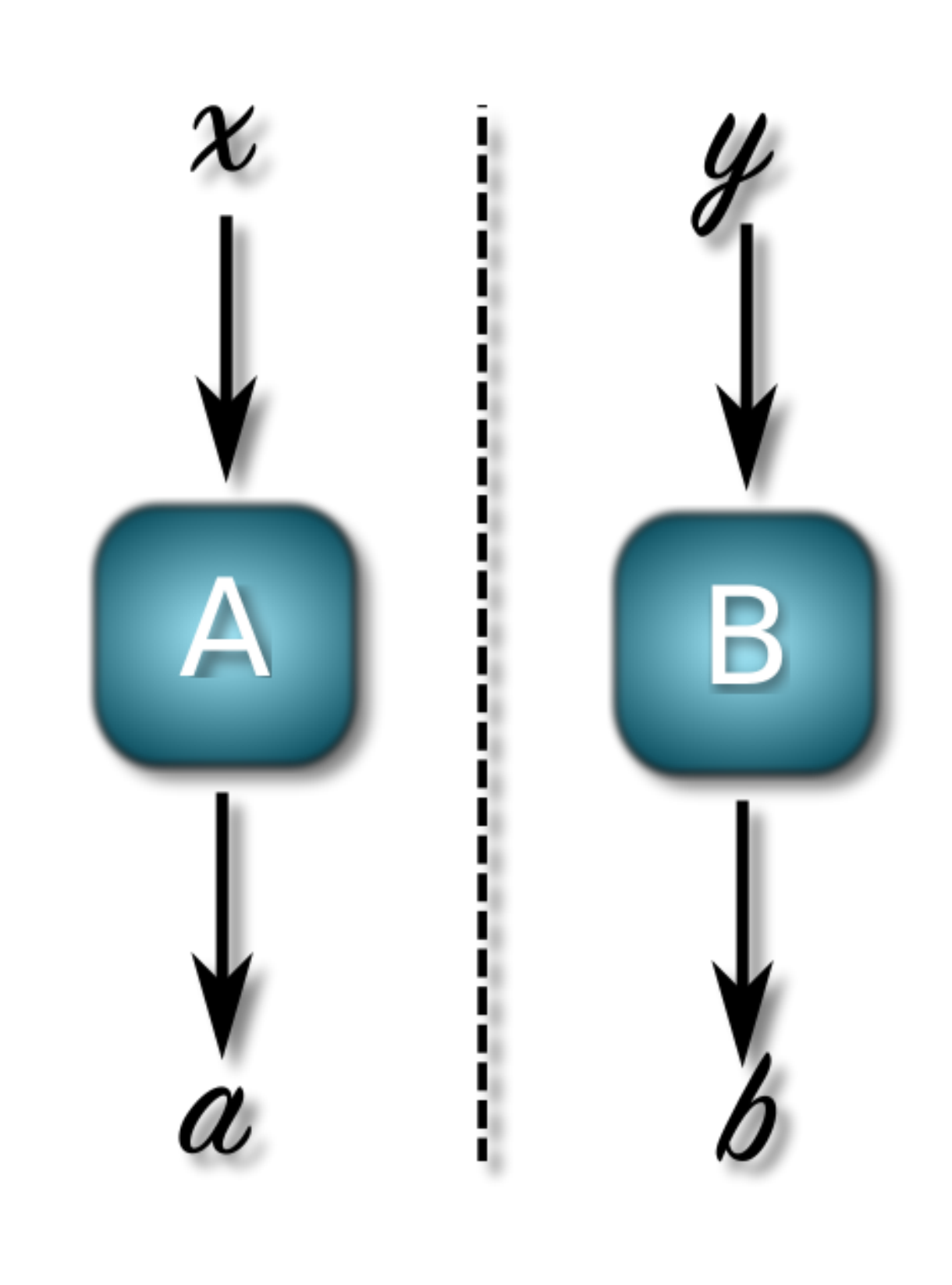}
		\caption[fig1]{Illustration of a bipartite scenario with two spacelike separated parties Alice (A) and Bob (B). Alice inputs $x$ and gets an output $a$ while Bob inputs $y$ and gets an output $b$. We consider the simplest Bell scenario where inputs and outputs are binary, i.e., $x,y,a,b \in \{0,1\}$.}	\label{fig1}
		
	\end{figure}

	We will focus throughout the paper on the simplest possible Bell scenario, two spacelike separated parties, Alice and Bob, that, upon receiving their parts of a joint physical system, can make different measurements (labeled by $X$ for Alice and $Y$ for Bob) obtaining the corresponding outcomes (labeled $A$ and $B$). All the empirically accessible data of such a simple experiment are encoded in the probability distribution $p(a,b\vert x,y)$  that represents the correlations shared between Alice and Bob (see Fig.~\ref{fig1}). Further, we consider that $x,y,a,b \in\{0,1\}$, i.e., we focus on the case of binary inputs and outputs, known as the CHSH scenario~\cite{chsh}. 
	
\begin{figure}[!ht]
		\includegraphics[scale=0.34]{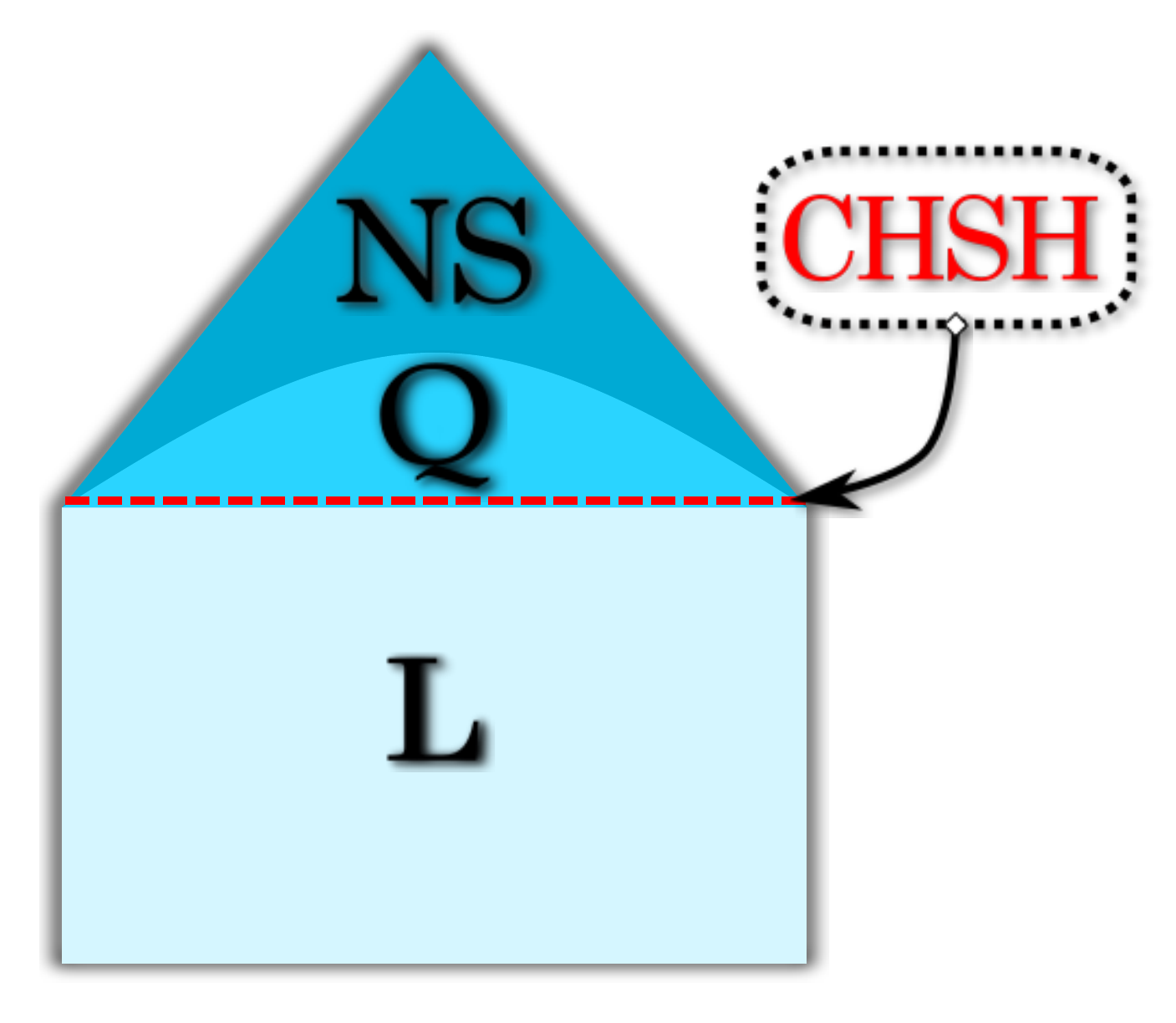}
		\caption[fig2]{An illustration of the strict inclusion relation $L\subsetneq Q \subsetneq$~NS in the bipartite two-input two-output scenario. The local face marked CHSH represents a hyperplane corresponding to the Bell-CHSH inequality.} 	\label{fig2}
		
	\end{figure}

	Within this context, three different sets of correlations are of remarkable importance. The set $L$ of \emph{local correlations} refers to probabilities $p(a,b \vert x,y)$ that can be reproduced by local hidden-variable models, that is,
	\begin{equation}
	p(a,b \vert x,y)= \sum_{\lambda} p(\lambda) p(a \vert x,\lambda)p(b \vert y,\lambda),
	\end{equation}
	where the measurement outcomes are local responses to their corresponding inputs and to the source of shared correlations represented by $\Lambda$. In turn, the quantum mechanical description based on Born's rule states that the set $Q$ of \emph{quantum correlations} is derived from
	\begin{equation}
	p(a,b \vert x,y)= \mathrm{Tr}\left[ \left(M^x_a \otimes M^y_b \right)\rho \right],
	\end{equation}
	where $\{M_a^x\}$ and $\{M_b^y\}$ represent measurement operators and $\rho$ a joint quantum state. Finally, we can define the set NS of \emph{no-signaling correlations} by the linear constraints,
	\begin{eqnarray}
	\label{NS}
	& & \sum_{b} p(a,b \vert x,y)= p(a \vert x)  = \sum_{b} p(a,b \vert x,y^{\prime}), \\
	& & \sum_{a} p(a,b \vert x,y)= p(b \vert y) =  \sum_{a} p(a,b \vert x^{\prime},y),
	\end{eqnarray}
	as saying that distant observers should not influence the measurement statistics of 
	each other (otherwise superluminal communication would be possible). A fundamental result in the geometry of Bell correlations states that $L\subsetneq Q \subsetneq NS$ (see Fig.~\ref{fig2}). The first strict inclusion follows from Bell's theorem~\cite{bell}, showing that there are nonlocal quantum correlations. The second strict inclusion follows from Ref.~\cite{pr} showing the existence of postquantum no-signaling correlations.
	
	In the CHSH scenario, the no-signaling correlations have been fully characterized in Ref.~\cite{bar}; the set forms an eight-dimensional polytope with 24 vertices. Out of these vertices, 16 are local whereas eight nonlocal. To represent all the $24$ vertices, we will use the notation $\delta_{r,s}$, meaning that 
	\[
	\delta_{r,s}= 
	\begin{cases}
	1,& \text{if } r=s \\
	0,              & \text{otherwise}.
	\end{cases}
	\] 
	
	The nonlocal vertices are the eight symmetries of the PR box and can be expressed in terms of parameters $\alpha,\beta,\gamma \in\{0,1\}$ as follows,
	\begin{equation}
	p^{\alpha,\beta,\gamma}_{\mathrm{PR}}(a,b \vert x,y)= \frac{1}{2} \delta_{a \oplus b, xy\oplus \alpha x \oplus \beta y \oplus \gamma}.
	\label{prbox}
	\end{equation}
	The ``canonical'' PR box corresponds to the parameters $(\alpha,\beta,\gamma)=(0,0,0)$. In turn, the 16 local vertices, which are merely all possible local deterministic probability distributions, can be written in terms of parameters  $\alpha_1,\alpha_2,\beta_1,\beta_2 \in\{0,1\}$ as follows,
	\begin{equation}
	p^{\alpha_1,\alpha_2,\beta_1,\beta_2,}_{\mathrm{L}}(a,b \vert x,y)=  \delta_{a , \alpha_1 x \oplus \alpha_2  }\delta_{b , \beta_1 y \oplus \beta_2}.
	\label{lbox}
	\end{equation}
	The local set $L$ is another polytope generated by the 16 local vertices, and it is an eight-dimensional subpolytope of the NS polytope. Moreover, as shown in Ref.~\cite{bar}, $L$ has exactly eight nontrivial faces which are in a one-to-one correspondence with the eight symmetries of the CHSH inequality \cite{chsh},
	\begin{eqnarray}
	\label{chsh0}
	& & p(0,0\vert 0,0)+p(0,0\vert 0,1)+p(0,0\vert 1,0) \\ \nonumber
	& & -p(0,0\vert 1,1) -p_A(0\vert 0)-p_B(0\vert 0) \leq 0 ,
	\end{eqnarray}
	where $p(a=0,b=0\vert x=0,y=0)=p(0,0\vert 0,0)$ (similarly to the other terms) and $ p_A(0\vert 0)=\sum_{b} p(a=0,b \vert x=0,y)$ is the marginal distribution of Alice [similarly for $p_B(0\vert 0)$].
	Each of the symmetries of the CHSH inequality is violated (maximally) by exactly one of the symmetries of the PR box. For example, the CHSH inequality (\ref{chsh0}) is violated maximally by the canonical PR box $p^{0,0,0}_{\mathrm{PR}}(a,b \vert x,y)=\frac{1}{2}\delta_{a\oplus b,xy}$, whereas none of the remaining seven PR boxes violates (\ref{chsh0}). Finally, from the strict inclusion relation $L\subsetneq Q \subsetneq$~NS, we see that the quantum set is an eight-dimensional set (a convex set though not a polytope). 
	
	By referring to the known symmetries, see the Ref.~\cite{Bierhorst}, of the no-signaling polytope in the CHSH scenario, without loss of generality, it is sufficient to consider nonlocal correlations in any one of the symmetric regions. Therefore, in this paper, we will focus on the nonlocal region defined by the convex hull of the canonical PR box (henceforth referred to simply as "the PR-box"), and the eight local vertices on the local face derived from "the CHSH inequality" given by condition~(\ref{chsh0}).
	
	\section{The nonlocal simplex and its faces} \label{nlsimplex}
	Let the set of $16$ joint probabilities, in the CHSH scenario,
	$\{p(a, b|x, y):a,b,x,y\in\{0,1\} \}$, 
	be represented as a vector $\vec{p}=(p_1,p_2,...,p_{16})\in \mathbb{R}^{16}$. We order the elements of this vector according to the following rule, $p(a,b \vert x,y)\rightarrow p_i$, where the index $i$ can be determined from $i=2^3c + 2^2x + 2^1y + a + 1 $, and $c=a\oplus b\oplus xy\oplus 1$. The symbol $\oplus$ represents addition modulo $2$. For example, in this ordering, the vector of probabilities corresponding to the PR box is $$\vec{p}_{PR}=(0,0,0,0,0,0,0,0,\frac{1}{2},\frac{1}{2},\frac{1}{2},\frac{1}{2},\frac{1}{2},\frac{1}{2},\frac{1}{2},\frac{1}{2}).$$
	
	We consider the eight probabilities $\{p_i : 1\leq i\leq 8\}$ as free variables. Then, the remaining eight probabilities $\{p_j : 9\leq j\leq 16\}$ can be written in terms of free variables by using the no-signaling and normalization conditions as follows,
	\begin{equation}
	p_j = p_l + p_m + p_n + \frac{1}{2}(1-\sum_{i=1}^8 p_i), 
	\label{dep-var}
	\end{equation}
	where the indices of free-variable probabilities $ p_l, p_m, p_n$ are given by
	\begin{eqnarray}
	l&=& 2^2x + 2(y\oplus 1) + a + 1, \\
	m&=& 2^2(x\oplus 1) + 2y +(a\oplus y\oplus 1)+ 1,\\
	n&=& 2^2(x\oplus 1) + 2(y\oplus 1) +(a\oplus y)+ 1 .
	\label{ind-free-var}
	\end{eqnarray}
	
	In terms of the joint probabilities which we consider as free variables, the CHSH inequality (\ref{chsh0}) takes the following form,
	\begin{equation}
	1-\sum_{i=1}^{8}p_i\leq 0.
	\label{chsh}
	\end{equation}  
	On the facet, of the local polytope $L$, corresponding to the CHSH inequality (\ref{chsh}), there are eight local vertices which we denote by $\{L_i: 1\leq i\leq 8\}$. The correspondence between these eight local vertices and the values of parameters $(\alpha_1,\alpha_2,\beta_1,\beta_2)$ is as follows:
	\begin{eqnarray}
	L_1\equiv(1,0,1,1),~~ L_2\equiv(1,1,1,0), \nonumber\\ L_3\equiv(0,0,1,0),~~L_4\equiv(0,1,1,1), \nonumber\\
	L_5\equiv(1,1,0,1),~~L_6\equiv(1,0,0,0),\nonumber\\L_7\equiv(0,0,0,0),~~L_8\equiv(0,1,0,1).
	\end{eqnarray}
	It turns out that for local vertex $L_i$, the free-variable probabilities $p_k=\delta_{ik}$, where $1\leq k\leq 8$.
	
	We consider the region which is a convex hull of the PR box and eight local vertices $\{L_i: 1\leq i\leq 8\}$. Henceforth we call this region a nonlocal region and denote it by $\mathbf{NL}$ (since all points in this region except those which are on the local face are nonlocal points). It turns out that this region forms an eight-dimensional simplex \cite{simplex}. To show this, we arrange all the vectors of the nine vertices of the $\mathbf{NL}$ polytope in a matrix. The first eight rows of this matrix correspond respectively to the eight local vertices $\{L_i: 1\leq i\leq 8\}$ and the ninth row represents the PR box; the resulting matrix is as follows:
	
	\[\left( {\begin{array}{ccccccccccccccccc}
		\vec{p}_{L_1}:&1 & 0 & 0 & 0 & 0 & 0 & 0 & 0 & 0 & 0 & 1 & 0 & 0 & 1 & 0 & 1\\
		\vec{p}_{L_2}:&0 & 1 & 0 & 0 & 0 & 0 & 0 & 0 & 0 & 0 & 0 & 1 & 1 & 0 & 1 & 0\\
		\vec{p}_{L_3}:&0 & 0 & 1 & 0 & 0 & 0 & 0 & 0 & 1 & 0 & 0 & 0 & 1 & 0 & 1 & 0\\
		\vec{p}_{L_4}:&0 & 0 & 0 & 1 & 0 & 0 & 0 & 0 & 0 & 1 & 0 & 0 & 0 & 1 & 0 & 1\\
		\vec{p}_{L_5}:&0 & 0 & 0 & 0 & 1 & 0 & 0 & 0 & 0 & 1 & 0 & 1 & 0 & 0 & 1 & 0\\
		\vec{p}_{L_6}:&0 & 0 & 0 & 0 & 0 & 1 & 0 & 0 & 1 & 0 & 1 & 0 & 0 & 0 & 0 & 1\\
		\vec{p}_{L_7}:&0 & 0 & 0 & 0 & 0 & 0 & 1 & 0 & 1 & 0 & 1 & 0 & 1 & 0 & 0 & 0\\
		\vec{p}_{L_8}:&0 & 0 & 0 & 0 & 0 & 0 & 0 & 1 & 0 & 1 & 0 & 1 & 0 & 1 & 0 & 0\\
		\vec{p}_{PR}:&0 & 0 & 0 & 0 & 0 & 0 & 0 & 0 & \frac{1}{2} &\frac{1}{2} &\frac{1}{2} &\frac{1}{2} &\frac{1}{2} & \frac{1}{2} &\frac{1}{2} &\frac{1}{2}\\
		\end{array} } \right).
	\] \label{matrix}
	
	It is now easy to see that the nine rows of the above given matrix are affinely independent, which means that on subtracting any one row from the remaining eight rows, the resulting eight vectors are linearly independent. Therefore, the region $\mathbf{NL}$ is simply an eight-dimensional simplex with nine vertices given by the PR box and eight local vertices $\{L_1,L_2,L_3,L_4,L_5,L_6,L_7,L_8\}$.

	Moving on, we now classify all the (proper) faces \cite{face} of the nonlocal simplex. Since we discuss faces of a simplex, it is easy to characterize all of them: Consider a $d$-dimensional simplex defined by a set of vertices $V=\{v_1,v_2,...,v_{d+1}\}$, then any nonempty proper subset $S\subsetneq V$ defines a (proper) face $F_S$ of the simplex, where the face $F_S$ is the convex hull of all points in $S$. The $\mathbf{NL}$ simplex has eight nonlocal facets of dimension seven (each defined by a convex hull of the PR box and seven local vertices) and one local facet of dimension seven (defined by a convex hull of eight local vertices). We note that there will be many nonlocal and local faces of dimension less than seven which are contained in the facets of highest dimension seven. Among these faces, all the nonlocal faces of the simplex can be determined as follows: Consider a nonempty subset $\{p_{i_{1}},...,p_{i_{k}}\}$ of the set of eight free-variable probabilities $\{p_1,p_2,...,p_8\}$, and then assigning the value zero to all probabilities of any such subset defines a nonlocal face. On the other hand, if we assign the value zero to any non-free-variable probability, then from Eq.~(\ref{dep-var}) it follows that  $(1-\sum_{i=1}^{8}p_i)\leq 0$, which means that the Bell-CHSH inequality~(\ref{chsh}) cannot be violated. Therefore, assigning the value zero to any non-free-variable probability will lead only to some local face(s). Since the interesting aspect is to analyze the nonlocal faces, in the remaining parts of this paper, we will deal only with nonlocal faces of different dimensions, i.e., faces which are derived by assigning the value zero to a number of free-variable probabilities. Thus, any nonempty subset $S$ of the set of all free-variable probabilities $\{p_i: 1\leq i \leq 8\}$ defines a nonlocal face $F_S$ if we set all probabilities in the subset $S$ to zero. The number $8-|S|$ in turn gives the dimension of the nonlocal face $F_{S}$. For example, if $S=\{p_i: 1\leq i \leq 8\}$, we get a face of dimension zero, and this face contains only one nonlocal point which is the PR box. At the other extreme, for example, $S=\{p_8\}$ defines a nonlocal face of maximal dimension seven which is simply a set of all points in the convex hull of the PR-box and seven local deterministic points $\{L1,L2,...,L7\}$.

	\section{Some examples of quantum correlations on nonlocal faces of the "$\mathbf{NL}$" simplex} \label{qHpoints}
	Let us denote set of all quantum points in the simplex $\mathbf{NL}$ as $Q_{NL}$. One interesting aspect is that there are nonlocal points in $Q_{NL}$ which live on some of the faces of the $\mathbf{NL}$ simplex, and in this section we give some such examples. Let the observable (or input) of Alice be denoted by $A_{\alpha}$ and that of Bob by $B_{\beta}$, where $\alpha, \beta \in \{0,1\}$. Now consider that, for parameters $a,x,y \in\{0,1\}$, a set of four joint probabilities satisfies the following four conditions,
	\begin{eqnarray}
	p(a,a\oplus xy\vert A_x,B_y)>0, \label{h1}\\
	p(i,a\oplus xy|A_{x\oplus 1},B_y)=0, \label{h2}\\
	p(a,j|A_x,B_{y\oplus 1})=0, \label{h3} \\
	p(i\oplus 1,j\oplus 1|A_{x\oplus 1},B_{y\oplus 1})=0, \label{h4}
	\end{eqnarray}
	where
	\begin{eqnarray}
	i=a\oplus y\oplus 1, \label{cond1}\\
	j=xy\oplus x\oplus a\oplus 1.\label{cond2}
	\end{eqnarray} 
	Then, from conditions (\ref{h1})-(\ref{h4}), for any choice of $i,j,a,x,y \in\{0,1\}$, one can give Hardy's argument \cite{hardy} showing that the joint probability distribution satisfying these four conditions must be nonlocal. The joint probability in condition (\ref{h1}), in turn, is referred to as the success probability of Hardy's argument. Hardy's nonlocality argument is as follows: Suppose that a probability distribution which satisfies conditions (\ref{h1})-(\ref{h4}) has a local (deterministic \cite{lhv}) hidden-variable model with hidden variables $\lambda \in \Lambda$. Then, condition (\ref{h1}) implies that there is at least one $\lambda_{*} \in \Lambda$ for which $A_x(\lambda_{*})=a$ and $B_y(\lambda_{*})=a\oplus xy$. Now Eqs.~(\ref{h2})~and~(\ref{h3}) respectively imply that for the hidden variable $\lambda_{*}$, $A_{x\oplus 1}(\lambda_{*})=i\oplus 1$ and $B_{y\oplus 1}(\lambda_{*})    =j\oplus 1$. However, Eq.~(\ref{h04}) shows that there cannot be any hidden variable $\lambda$ such that $A_{x\oplus 1}(\lambda)=i\oplus 1$ and $B_{y\oplus 1}(\lambda)=j\oplus 1$, which is a contradiction. Therefore, any probability distribution satisfying conditions (\ref{h1})-(\ref{h4}) cannot have a local hidden-variable model and hence it must be nonlocal.
	
	The choice of condition on $i$ and $j$ in Eqs.~(\ref{cond1})~and~(\ref{cond2}) ensures that the joint probabilities appearing in Eqs.(\ref{h2})-(\ref{h4}) belong to the set of free-variable probabilities. Notice that the nonzero probability in the condition~(\ref{h1}) does not belong to the set of free-variable probabilities. To sum up, conditions (\ref{h1})-(\ref{cond2}) basically give all Hardy nonlocal points contained in the $\mathbf{NL}$ simplex that we consider. For instance, the PR box (a vertex of the $\mathbf{NL}$ simplex and a Hardy nonlocal point) satisfies all the conditions (\ref{h1})-(\ref{cond2}).
	
	We know that Hardy nonlocality arguments have quantum solutions \cite{hardy,jordan,kar}. In Ref.~\cite{jordan} it was shown that given any set of qubit measurements (such that measurements of each party do not commute) one can always find a two-qubit entangled state leading to Hardy's nonlocality argument. In turn, it was shown in  Ref.~\cite{kar} that for a given set of measurements there is a unique pure state achieving such correlations. Therefore, since all possible Hardy nonlocality arguments given by conditions (\ref{h1})-(\ref{cond2}) have quantum solutions, we know that every seven-dimensional nonlocal facet of the $\mathbf{NL}$ simplex contains nonlocal quantum correlations.
	
	\section{Characterization of all nonlocal faces of the $\mathbf{NL}$ simplex} \label{NLfaces}
	Another interesting aspect of nonlocal faces of the $\mathbf{NL}$ simplex is that some of these faces are \emph{quantum voids}, i.e., all the nonlocal points on these faces are postquantum. In what follows, first we prove the existence of quantum voids of maximal possible dimension, and then we use the result to characterize all nonlocal faces of any dimension into two categories: (i) quantum voids and (ii) not quantum voids. 
	\subsection{Six-dimensional quantum voids on nonlocal faces}          
	We first show that the nonlocal simplex contains nonlocal faces of maximal possible dimension six in which all nonlocal correlations (points with a nonzero weight for the PR box) are postquantum. 
	
	Consider six-dimensional nonlocal faces defined by assigning the value zero to the following two free-variable probabilities,
	\begin{eqnarray}
	p(a\oplus 1, a\oplus xy | x,y)=0, \label{6dvoid-eq1}\\
	p(a\oplus 1, a\oplus xy\oplus x| x,y\oplus 1)=0, \label{6dvoid-eq2}
	\end{eqnarray}
	where $a,x,y\in \{0,1\}$.\\
	
	\emph{Proposition~1.}~Six-dimensional nonlocal faces of the $\mathbf{NL}$ simplex defined by Eqs.~(\ref{6dvoid-eq1}) and (\ref{6dvoid-eq2}) are quantum voids.
	\begin{proof}
		Let the two supporting hyperplanes, $p(a\oplus 1, a\oplus xy | x,y)=0$  and $p(a\oplus 1, a\oplus xy\oplus x| x,y\oplus 1)=0$, of the eight-dimensional quantum region $Q_{NL}$ be denoted respectively by $H_{1}$ and $H_{2}$. Consider the section of the quantum region defined by ${Q}^{12}_{NL}=Q_{NL}\cap H_{1}\cap H_{2}$. Since $Q_{NL}$ is a convex set, all the extremal points of $Q^{12}_{NL}$ are also extremal points of $Q_{NL}$, which in turn are extremal points of the quantum set $Q$. From the results in Refs.~\cite{mas1,mas2}, it then follows that all the extremal points of $Q^{12}_{NL}$ can be achieved via projective measurements on two-qubit pure states. By optimizing over all such measurements and states we will show that all correlations in $Q^{12}_{NL}$ are local, thus proving the quantum void.
		
		Consider quantum correlations generated by a two-qubit pure state $|\Psi\rangle$, the two projective measurements performed by Alice given by the orthonormal basis $\{\ket{\alpha^x_0},\ket{\alpha^x_1}\}$ and $\{\ket{\alpha^{(x\oplus 1)}_0},\ket{\alpha^{(x\oplus 1)}_1}\}$, and similarly for Bob given by $\{\ket{\beta^y_0},\ket{\beta^y_1}\}$ and $\{\ket{\beta^{(y\oplus 1)}_0},\ket{\beta^{(y\oplus 1)}_1}\}$. In order to represent the state $|\Psi\rangle$, we choose an orthonormal basis given by $\{\ket{\alpha^x_0} \otimes \ket{\beta^y_0},\ket{\alpha^x_0} \otimes \ket{\beta^y_1},\ket{\alpha^x_1} \otimes \ket{\beta^y_0},\ket{\alpha^x_1} \otimes \ket{\beta^y_1}\}$. The first constraint Eq.~(\ref{6dvoid-eq1}) implies that
		
		\begin{equation}
		\langle\alpha^x_{a\oplus 1}\beta^y_{a\oplus xy}|\Psi\rangle=0, \label{Eq(i)}
		\end{equation}
		
		and the second constraint Eq.~(\ref{6dvoid-eq2}) implies that
		\begin{equation}
		\langle\alpha^x_{a\oplus 1}\beta^{y\oplus 1}_{a\oplus xy\oplus x}|\Psi\rangle=0. \label{Eq(ii)}
		\end{equation}
		
		Since $\{\ket{\beta^y_0},\ket{\beta^y_1}\}$ defines a basis of the second party, we can write
		$|\beta^{y\oplus 1}_{a\oplus xy\oplus x}\rangle=c_0\ket{\beta^y_{a\oplus xy}}+c_1\ket{\beta^y_{a\oplus xy\oplus 1}}.$ Therefore, Eq.~(\ref{Eq(ii)}) can now be written as
		\begin{equation}
		c_0^{\ast}\bra{\alpha^x_{a\oplus 1}\beta^y_{a\oplus xy}}\Psi\rangle+c_1^{\ast}\bra{\alpha^x_{a\oplus 1}\beta^y_{a\oplus xy\oplus 1}}\Psi\rangle=0. \label{Eq(iii)}
		\end{equation}
		Then, from Eqs.(\ref{Eq(i)}) and (\ref{Eq(iii)}), it follows that either $c_1=0$ (meaning that Bob's measurements commute and thus can only lead to local correlations), or that $\langle\alpha^x_{a\oplus 1}\beta^y_{a\oplus xy}|\Psi\rangle=0$ and $\bra{\alpha^x_{a\oplus 1}\beta^y_{a\oplus xy\oplus 1}}\Psi\rangle=0$, implying  $\ket{\Psi}=|\alpha^x_{a}\rangle \otimes (k_0 |\beta^y_{a\oplus xy}\rangle + k_1 |\beta^y_{a\oplus xy \oplus 1}\rangle )$, which is a separable state (again, only generating local correlations). 
	\end{proof}
	
	Finally, we note that replacing the second constraint, i.e., Eq.~(\ref{6dvoid-eq2}), by the following equation,
	\begin{equation}
	p(a\oplus 1\oplus y, a\oplus xy| x\oplus 1,y)=0 \label{6dvoid-eq2b},
	\end{equation}
	we get another set of six-dimensional quantum voids (the proof runs similar to that for Proposition~1).
	
	\subsection{All quantum voids of the $\mathbf{NL}$ simplex}
	\begin{figure}[h]
		\includegraphics[scale=0.28]{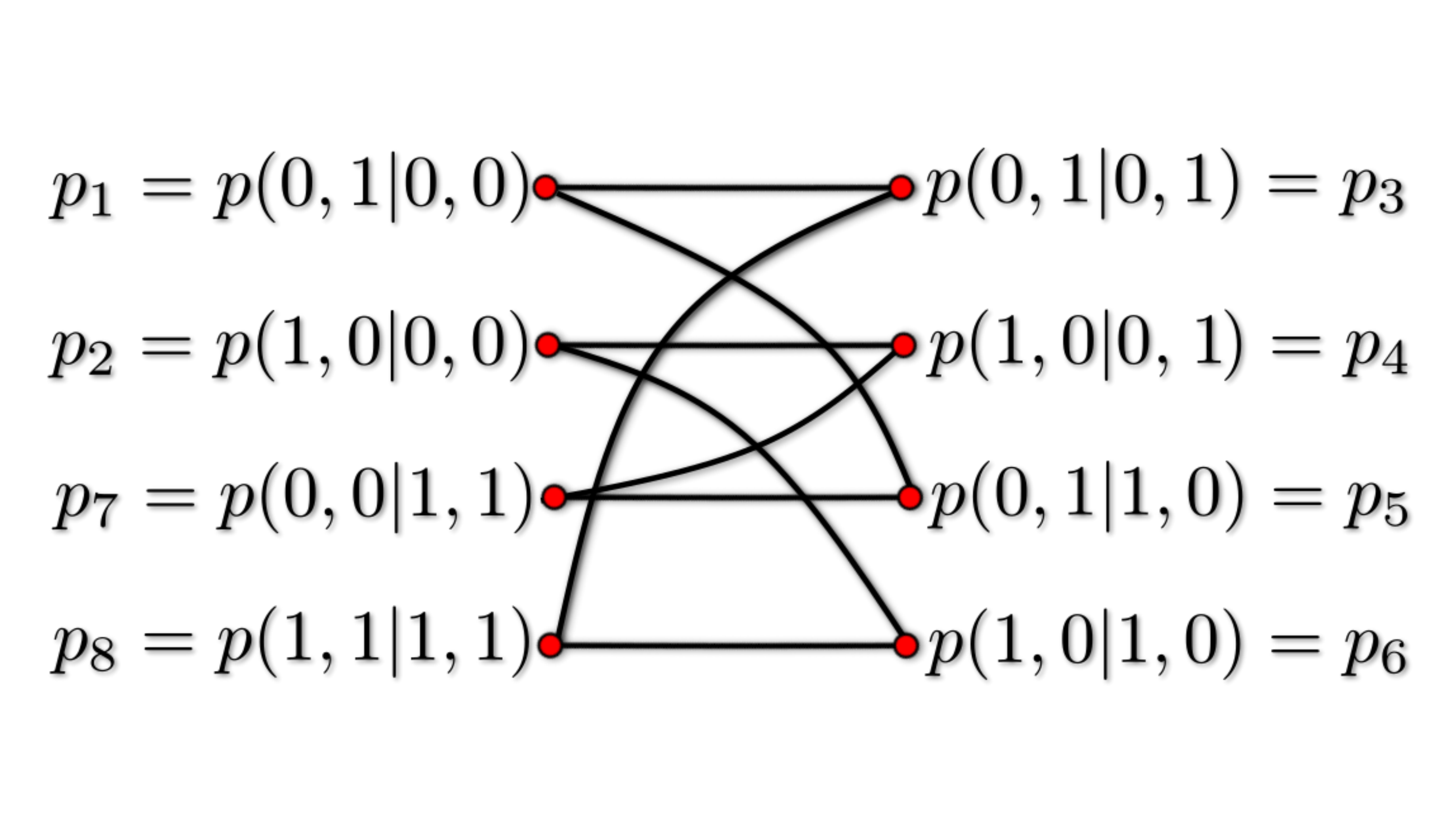}
		\caption[fig3]{A bipartite graph with all free-variable probabilities as nodes. An edge between two nodes means the two probabilities (when assigned the value zero) define a six-dimensional quantum void. }	\label{fig3}
	\end{figure}
	
	To characterize all the nonlocal faces of the $\mathbf{NL}$ simplex into two categories, (i) quantum voids, and (ii) not quantum voids, first, we use the results that we have derived for six-dimensional quantum voids. All these results are summarized in Fig.~\ref{fig3}, which is a bipartite graph with the eight free-variable probabilities as nodes, and an edge connecting two nodes means that assigning value zero to the two probabilities corresponding to the two nodes leads to a six-dimensional quantum void.

\begin{figure}[t]
		\includegraphics[scale=0.30]{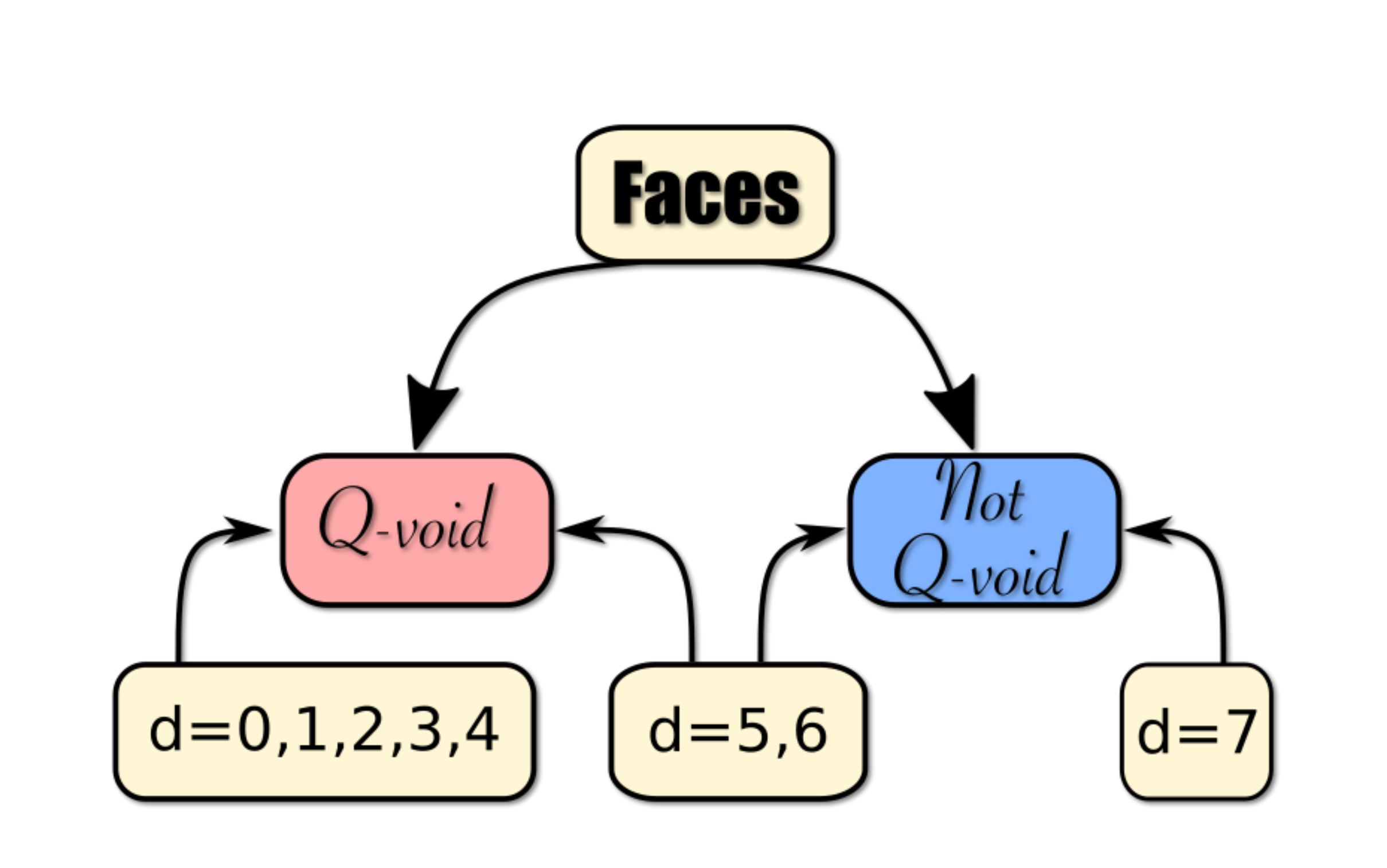}
		\caption[fig4]{An illustration of the relation between different dimensional faces of the $\mathbf{NL}$ simplex, and the two categories: (i) Q-void and (ii) not Q-void. Here, one can see that all faces of dimension $d\leq4$ are Q-voids; five- and six-dimensional faces can be of both types, Q-voids and not Q-voids; and all seven-dimensional faces are not Q-voids.}	\label{fig4}
		
	\end{figure}

	From Fig.~\ref{fig3} one can find all possible quantum voids, except for two of them. Basically, this follows first by noticing that all the lower-dimensional faces of the six-dimensional voids are also quantum voids. Next, on considering all the remaining nonlocal faces, i.e., those defined by subsets of free-variable probabilities such that according to Fig.~\ref{fig3} there is no edge between any two points of these subsets, we find that, with two exceptions, all such subsets define faces containing quantum nonlocal points (see the Appendix for details). In Fig.~\ref{fig3}, let us name the set of points in column~$1$ as $S_1$ and that in the column~$2$ as $S_2$. The two quantum voids which do not follow from Fig.~\ref{fig3} are the four-dimensional nonlocal faces defined by the subsets $S_1=\{p_1,p_2,p_7,p_8\}$ and $S_2=\{p_3,p_4,p_5,p_6\}$ (see the Appendix for a proof). We summarize the information about the nonlocal faces of given dimensions, which can be a quantum void or not a quantum void, in Fig.~\ref{fig4}. Here, we note that some of the four-dimensional quantum voids have been derived before in Ref.~\cite{fritz}, which is study of Hardy's nonlocality paradox for the set of NS correlations.
	
	\section{Certain class of Convex sets in the nonlocal simplex} \label{convex}
	After characterizing all the nonlocal faces into two classes, (i) faces that are quantum voids and (ii) faces that are not quantum voids, we study the two classes in light of some known physical principles and the almost quantum set $Q^{(1+ab)}$. It is known that all levels of the NPA outer approximations of $Q$ are convex sets \cite{lang}, thus, a set of correlations respecting the macroscopic locality principle (level 1 of the NPA hierarchy), and the set of almost quantum correlations $Q^{(1+ab)}$ (level $1+ab$ of the NPA hierarchy), are convex sets. On the other hand, it is known that one of the known necessary conditions for respecting the information causality principle is Uffink's inequality~\cite{uffink}, which again generates a convex set \cite{allcock2}. These observations lead us to derive a general result which will cover all the above-mentioned types of convex sets. We simply denote such sets by $Q^{(k)}$, where the superscript $k$ will be replaced by a particular convex set as required.
	
	So we consider convex outer approximations $Q^{(k)}$ of the quantum set $Q$ restricted to the nonlocal simplex $\mathbf{NL}$. As the intersection of two convex sets is convex, the resulting restricted set $Q^{(k)}_{NL}=Q^{(k)} \cap NL$ is also convex.  We will use the notation $ch$ for the convex hull of a set of points. Now consider a representation of \emph{the nonlocal boundary}, denoted by $\partial Q^{(k)}_{NL}$, as a function from the set of points on the local face to some real number, i.e., $\partial Q^{(k)}_{NL}:ch(\{L_1,...,L_8\})\rightarrow \mathbb{R}$. We define such a function as follows: We consider a local point $x\in L=ch(\{L_1,...,L_8\})$ and join it with a point representing the PR box; set of all points on this line segment can be expressed as $\mu PR +(1-\mu)L_x$ where $0\leq \mu \leq 1$. Now we define $\partial Q^{(k)}_{NL}(x)=\mu^{(k)}_{*}$ so that $\mu^{(k)}_{*}$ is the maximum value of $\mu^{(k)}$ on the considered line segment. The region $Q^{(k)}_{NL}$  is the hypograph of the function  $\partial Q^{(k)}_{NL}(x)=\mu^{(k)}_{*}$; moreover, since  $Q^{(k)}_{NL}$ is convex this implies that the function $\partial Q^{(k)}_{NL}(x)=\mu^{(k)}_{*}$ is a concave function~\cite{hypograph}. Note that $\partial Q^{(k)}_{NL}(x)$ is a non-negative function.
	
	Since we are studying the nonlocal boundary of $Q$ on faces of the $\mathbf{NL}$ simplex, we define the restrictions to a convex set $Q^{(k)}$ in the lower-dimension simplexes $ch(\{PR,L_{i_{1}},...,L_{i_{d}}\})$, where $S_L=\{L_{i_{1}},...,L_{i_{d}}\}\subsetneq \{L_1,L_2,...,L_8\}$ is a set of $d$ local deterministic points. Then $\partial Q^{(k)}_{NL}(x)=\mu^{(k)}_{*} ~\forall~ x\in ch(S_L)$ and it is a non-negative concave function.\\
	
	\emph{Proposition~2.} Consider some interior point $x_{0}$ of the domain $ch(S_L)$, then $\partial Q^{(k)}_{NL}(x_{0})=0$ if and only if $\partial Q^{(k)}_{NL}(x)=0$ for all points $x \in ch(S_L)\backslash\{x_0\}$.
	\begin{proof}
		\emph{Only if part}, suppose that $\partial Q^{(k)}_{NL}(x_{0})=0$ and $\partial Q^{(k)}_{NL}(x)>0$ for some $x \in ch(S_L)$. Then, given that $x_{0}$ is an interior point, there exists a point $y \in ch(S_L)$ such that $x_{0}=\lambda x + (1-\lambda)y$, where $\lambda \in (0,1)$. Since the function $\partial Q^{(k)}_{NL}(\cdot)$ is concave, $\partial Q^{(k)}_{NL}(x_{0})\geq \lambda \partial Q^{(k)}_{NL}(x) +(1-\lambda)\partial Q^{(k)}_{NL}(y)>0$, which contradicts our initial assumption that $\partial Q^{(k)}_{NL}(x_0)=0$. 
		
		Next, for \emph{if part}, suppose that $\partial Q^{(k)}_{NL}(x_{0})>0$ and $\partial Q^{(k)}_{NL}(x)=0$ for all $x \in ch(S_L)\backslash\{x_0\}$. Given that $x_0$ is an interior point, there exist two distinct points $\{y,z\} \in ch(S_L)\backslash \{x_0\}$ such that $y=\lambda x_0+(1-\lambda) z$, where $\lambda \in (0,1)$. Then, from concavity of the function $\partial Q^{(k)}_{NL}(\cdot)$, we get $\partial Q^{(k)}_{NL}(y)\geq \lambda\partial Q^{(k)}_{NL}(x_0) +(1-\lambda)\partial Q^{(k)}_{NL}(z)$, implying that $\partial Q^{(k)}_{NL}(x_0)=0$; this contradicts our initial assumption $\partial Q^{(k)}_{NL}(x_0)>0$.
	\end{proof}

	\section{Physical principle(s) and quantum voids} \label{pp&qvoids}
	
	In this section we study the quantum voids in light of some known physical principles and the almost quantum set $Q^{(1+ab)}$. For this purpose, the result about the class of convex sets derived in the previous section will be used.  Since the exact analytical expressions for the necessary condition for respecting the information causality principle, i.e., Uffink's inequality \cite{uffink}, and the macroscopic locality principle \cite{ml} are known, we first checked the reproducibility of the quantum voids with the help of these two principles. Next, as required, we check the reproducibility of some of the quantum voids by the set of almost quantum correlations $Q^{(1+ab)}$ \cite{aq}, which in turn empowers us to draw a general conclusion about the reproducibility of quantum voids from all known physical principles \cite{ntcc,lnc,ic,ml,lo}. Before presenting our results below, first, in the following subsections, we give a brief account of the information causality principle, the macroscopic locality principle, and the set of almost quantum correlations $Q^{(1+ab)}$.

	\subsection{Information causality principle} 
	The information causality principle \cite{ic,ic-review} can be formulated quantitatively through an information-processing game played
	between two parties, Alice and Bob. Alice receives a randomly generated $N$-bit string $\vec{x} = (x_0,x_1, . . . ,x_{N-1})$, and Bob is asked to guess Alice's $i$th bit, where $i$ is a random question from the set $\{ 0,1,2, . . . ,N-1 \}$. Alice is allowed to send an $M$-bit message ($M <N$). Say Bob’s answer is $\beta_i$. Then, the information that Bob can potentially acquire about the bit $x_i$ of Alice is given by the Shannon mutual information $I (x_i : \beta_i )$. The statement of the information causality principle is that the total potential information about Alice’s bit string $\vec{x}$ accessible to Bob cannot exceed the quantity of the message he received from Alice, i.e., $\mathbf{I}=\sum_{i=1}^{N} I (x_i : \beta_i )\leq H(M)$ []$H(M)$ being the Shannon entropy of the message]. It is known that all quantum correlations respect the information causality principle. For the bipartite two-input and two-output scenario it was shown that a necessary condition for respecting the information causality principle turns out to be Uffink's inequality~\cite{uffink} (one of the first examples of a polynomial Bell inequality developed further in a recent work \cite{rafael2}). Uffink's inequality is as follows,
	\begin{equation}
	(C_{00}+ C_{10})^2 +(C_{01}- C_{11})^2 \leq 4, 
	\label{iccond}
	\end{equation} 
	where $C_{xy}=p(a\oplus b=0\vert x,y)-p(a\oplus b=1\vert x,y)$. Therefore, any violation of Uffink's inequality implies that the information causality principle is violated. A condition that is both necessary and sufficient for respecting the information causality principle is not known to date. However, a generalization of this principle \cite{generalized-ic} leads to tighter conditions which enables detecting some postquantum nonlocal correlations where Uffink's inequality fails. On the other hand, it is known that the set of correlations defined by Uffink's inequality is a convex set but it is not closed under wirings~\cite{allcock2}; this implies that some correlations which do not violate Uffink's inequality can do so \cite{allcock2,kunkri} by nonlocality distillation \cite{forster,brunner,hoyer}. We will focus on testing the necessary condition for respecting the information causality principle, i.e., Uffink's inequality.

	\subsection{Macroscopic locality principle} 
	The principle of macroscopic locality, in a bipartite scenario, states that if $N$ independent pairs of particles are emitted from sources such that the information about which source emitted which pair is lost, and only a coarse-grained measurement is possible on a bunch of $N$ particles at both ends, then if the number $N$ is very large (i.e., $N\rightarrow \infty$), the measurement statistics of such a coarse-grained (macroscopic) experiment cannot lead to a violation of any Bell inequality, i.e., the coarse grained statistics will have a local hidden-variable model (see Ref.~\cite{ml} for more details). Quantum correlations respect the macroscopic locality principle, however, it is not true for all NS correlations, for example, the PR box does not respect this principle. In Ref.~\cite{npa1} it was shown that, in a bipartite, two-input, and two-output scenario, the set of correlations generated by the level~$1$ of the NPA hierarchy $Q^{(1)}$ is exactly those correlations which respect the macroscopic locality principle. Moreover, this set can be exactly identified by an analytical condition, which is both necessary and  sufficient for respecting this principle. This condition is as follows: If $C_x\neq 1$ and $C_y\neq 1$ for all $x,y \in\{0,1\}$,
	
	\begin{equation}
	\left\vert \sum_{x,y=0}^1 (-1)^{xy}~\mbox{sin}^{-1} D_{xy}~\right\vert \leq \pi,
	\label{mlcond}
	\end{equation}
	where $D_{xy}=\frac{C_{xy}-C_{x}C{y}}{\sqrt{(1-C_x^2)(1-C_y^2)}}$, $C_{xy}=\sum_{a=b}p(a,b\vert x,y)-\sum_{a\neq b}p(a,b\vert x,y)$, $C_x=\sum_b [p(0,b\vert x,0)-p(1,b\vert x,0)]$, $C_y=\sum_a [p(a,0\vert 0,y)-p(a,1\vert 0,y)]$.
	Otherwise, the correlations are local-deterministic and they respect the macroscopic locality principle.

	\subsection{Set of almost quantum correlations}
	The set of almost quantum correlation $Q^{(1+ab)}$ was proposed in Ref.~\cite{aq} as a possible set of correlations which may possibly exist in nature, however, a recent study \cite{nrh} by some of the same authors has demonstrated that this now seems very unlikely. The set $Q^{(1+ab)}$ corresponds to the $1+ab$ level of the NPA hierarchy, and hence it is an outer approximation to the quantum set $Q$. One of the very interesting properties of the set $Q^{(1+ab)}$, as shown in Ref.~\cite{aq}, is that correlations which are postquantum but live inside $Q^{(1+ab)}$ cannot be reproduced (with our current knowledge) by all the so far proposed physical principles~\cite{aq}.

	\subsection{Reproducibility of quantum voids by physical principles}
	For testing the strength of the outer approximations $Q^{(k)}$ in reproducing any $d$-dimensional quantum void, note that, empowered by Proposition~2, it is now enough to compute the values of $\mu^{(k)}_{*}$ on any one line segment joining the PR box with some interior local point $x_{int}\in ch(S_L)$. We suitably choose $d$ local deterministic vertices such that $S_L=\{L_{i_{1}},...,L_{i_{d}}\}$ defines the local face of a $d$-dimensional quantum void. Thus, for studying these voids it is enough to consider an equal mixture (center point) of all local deterministic points, i.e., $L_{c}=\frac{1}{d}(L_{i_{1}}+...+L_{i_{d}})$, and then consider the line segment joining the PR box to the local point $L_{c}$. On this line segment, if $\mu^{(k)}_{*}=0$, we conclude that the considered quantum void can be reproduced by $Q^{(k)}$, otherwise it is not.
	
	On applying the necessary condition for respecting information causality principle (\ref{iccond}), i.e., the Uffink's inequality, and the macroscopic locality principle~(\ref{mlcond}), we get examples of the following three possibilities, quantum voids reproducible by (i) both information causality and macroscopic locality, (ii) macroscopic locality but not by Uffink's inequality, and (iii) neither by Uffink's inequality nor by macroscopic locality. Examples of quantum voids for all the three cases exist up to four-dimensional quantum voids. However, for all five- and six-dimensional quantum voids neither Uffink's inequality nor macroscopic locality could reproduce any of the quantum voids. In what follows, we give a few examples.
	
	\emph{One-dimensional quantum voids.} All these quantum voids are reproducible both by Uffink's and macroscopic locality conditions. For example, on considering $\mu PR+ (1-\mu) L_1$, both the information causality and macroscopic locality principles are violated for all $0<\mu \leq 1$. 
	
	\emph{Two-dimensional quantum voids.} All these quantum voids are reproducible by the macroscopic locality principle. Some of these voids can be reproduced by macroscopic locality but not by Uffink's inequality, for example, $ch\{PR,L_1,L_3\}$; on the other hand, some of these voids can be reproduced both by macroscopic locality and Uffink's inequality, for example, $ch\{PR,L_1,L_5\}$.

	\emph{Three-dimensional quantum voids.} Here, we have examples of all three types. An example of a void reproducible by both information causality and macroscopic locality is $ch\{PR,L_1,L_5,L_6\}$, an example of a void reproducible by macroscopic locality but not by information causality is $ch\{PR,L_1,L_3,L_4\}$, and an example of a void which cannot be reproduced either by Uffink's inequality or by macroscopic locality is $ch\{PR,L_1,L_4,L_5\}$.\
	
	\emph{Four-dimensional quantum voids.} Here, also we have examples of all three types. An example of a void reproducible by both information causality and macroscopic locality is $ch\{PR,L_1,L_2,L_5,L_6\}$, an example of a void reproducible by macroscopic locality but not by Uffink's inequality is $ch\{PR,L_1,L_2,L_3,L_4\}$, and an example of a void which cannot be reproduced either by Uffink's inequality or by macroscopic locality is $ch\{PR,L_1,L_3,L_4,L_5\}$
	
	\emph{Five- and six-dimensional quantum voids.} For all the five- and six-dimensional quantum voids, we find that none of these voids can be reproduced by Uffink's inequality or macroscopic locality conditions. Therefore we checked if some of these quantum voids can be reproduced by the set of almost quantum correlations $Q^{(1+ab)}$. We checked the $1+ab$ level of the NPA hierarchy and find that none of the six-dimensional quantum voids can be reproduced by $Q^{(1+ab)}$. For example, by considering the six-dimensional quantum void $ch(PR,L_1,L_2,L_3,L_4,L_5,L_7)$ we find the maximum value $\mu_{*}^{Q^{(1+ab)}}$, for the correlations on the line segment $\mu PR+\frac{1-\mu}{6}(L_1+L_2+L_3+L_4+L_5+L_7)$, over the set of almost quantum correlations $Q^{(1+ab)}$. For this we wrote a program in MATLAB, calling the function NPAHierarchy($\vec{p},1+ab$) from QETLAB~\cite{qetlab}, and computed the maximum value of $\mu_{*}^{Q^{(1+ab)}}$ for the set of almost quantum correlations. We find that  $\mu_{*}^{Q^{(1+ab)}}\simeq0.00094$, which is a strictly positive number. Then it follows from Proposition~$2$ that the considered six-dimensional quantum void cannot be reproduced by the set of almost quantum correlation $Q^{(1+ab)}$. We get similar results by considering all the other six-dimensional voids. Our result shows that, with our actual knowledge, the six-dimensional quantum voids cannot be reproduced  by any of the so far proposed physical principles.

	\section{Physical principles and nonlocal quantum points on the faces}\label{pp&qnlfaces}
	There are many quantum nonlocal points on the faces which are not quantum voids. On these faces, some of the known points on the boundary of the quantum set are the maximally nonlocal Hardy points \cite{rabelo,siba}, which give a maximum success probability of $(5\sqrt{5}-11)/2$ in quantum mechanics; we recall that the joint probability appearing in the condition (\ref{h1}) of the set of conditions leading to Hardy nonlocal points is termed as the success probability of Hardy's nonlocality argument. Interestingly, the quantum state and measurements leading to these points can be self-tested \cite{rabelo} so the resulting probability distribution is extremal points of the quantum set. Being extremal points, these can be generated by projective measurements on two-qubit pure entangled states \cite{mas1,mas2}. Then it is natural to check the reproducibility of these points by applying the known physical principles. It has been shown that these points cannot be reproduced by applying the physical principles such as the known condition for information causality \cite{ic-hardy1}, macroscopic locality \cite{ml-hardy}, and local orthogonality \cite{lo}. We will show here that even the set of almost quantum correlations $Q^{(1+ab)}$ \cite{aq} cannot reproduce the maximally (quantum) nonlocal Hardy points.
	
	We consider here only one of Hardy's nonlocality arguments given by substituting $a=x=y=0$ and $i=j=1$ in conditions (\ref{h1})-(\ref{h4}), which is as follows:
	
	\begin{eqnarray}
	p(0,0|A_0,B_0)>0, \label{h01}\\
	p(1,0|A_1,B_0)=0, \label{h02}\\
	p(0,1|A_0,B_1)=0, \label{h03} \\
	p(0,0|A_1, B_1)=0. \label{h04}
	\end{eqnarray}
	
	The Hardy points given by conditions (\ref{h01})-(\ref{h04}) live on the five-dimensional face of the $\mathbf{NL}$ simplex defined by the convex hull of the PR box and five local deterministic points $\{L_1,L_2,L_4,L_5,L_8\}$. It turns out that the maximally nonlocal Hardy point in the quantum set $Q$ can be obtained by two parties, Alice and Bob, on sharing the two-qubit quantum state
	\begin{equation}
	\vert \Psi_H\rangle=c_{00}\vert 00\rangle+c_{01}\vert 01\rangle+c_{10}\vert 10\rangle+c_{11}\vert 11\rangle,
	\end{equation}
	where $c_{00}=-\sqrt{\frac{5\sqrt{5}-11}{2}}$, $c_{01}=c_{10}=\frac{-3+\sqrt{5}}{2}$, $c_{11}=\sqrt{\frac{\sqrt{5}-1}{2}}$, and performing the projective measurements 
	
	\begin{eqnarray}
	A_0=B_0=\vert 0\rangle \langle 0\vert - \vert 1\rangle \langle 1\vert,\\
	A_1=B_1=\vert v\rangle \langle v\vert - \vert \bar{v}\rangle \langle \bar{v}\vert,
	\end{eqnarray}
	where $\vert v\rangle =\cos{\alpha}\vert 0\rangle+\sin{\alpha}\vert 1\rangle$,  $\vert \bar{v}\rangle =\sin{\alpha}\vert 0\rangle-\cos{\alpha}\vert 1\rangle$, and $\alpha = 2~\mbox{tan}^{-1}\left(\sqrt{-2+\sqrt{5}}\right)$.

	Then, the corresponding quantum probability distribution can be computed from the Born rule,
	
	\begin{equation}
	p(a,b\vert x,y)=\mbox{Tr}\left[\vert\Psi_H\rangle\langle\Psi_H\vert ~\Pi_{a}^{A_{x}}\otimes\Pi_{b}^{B_{y}}\right]
	\end{equation}
	where $\Pi_{a}^{A_{x}}$ is the projector corresponding to outcome $a$ of  measurement $A_{x}$, and $\Pi_{b}^{B_{y}}$ is the projector corresponding to outcome $b$ of measurement $B_{y}$. The resulting Hardy probability distribution is given in Table \ref{table1}.

	\begin{table}[h!]
		\begin{center}
			\caption{The maximal Hardy nonlocal probability distribution $P(ab\vert xy)$ in quantum mechanics, corresponding to the Hardy nonlocality argument given by conditions (\ref{h01})-(\ref{h04}).}
			\label{table1}
			\begin{tabular}{|l|*{4}{c|}}\hline
				\backslashbox{$x~y$}{$a~b$}
				&\makebox[3.5em]{$0~0$}&\makebox[3.5em]{$0~1$}&\makebox[3.5em]{$1~0$}&\makebox[3.5em]{$1~1$}\\\hline
				$0~0$ &$\frac{5\sqrt{5}-11}{2}$&$\frac{7-3\sqrt{5}}{2}$&$\frac{7-3\sqrt{5}}{2}$&$\frac{-1+\sqrt{5}}{2}$\\\hline
				$0~1$ &$-2+\sqrt{5}$&$0$&$\frac{7-3\sqrt{5}}{2}$&$\frac{-1+\sqrt{5}}{2}$\\\hline
				$1~0$ &$-2+\sqrt{5}$&$\frac{7-3\sqrt{5}}{2}$&$0$&$\frac{-1+\sqrt{5}}{2}$\\\hline
				$1~1$ &$0$&$\frac{3-\sqrt{5}}{2}$&$\frac{3-\sqrt{5}}{2}$&$-2+\sqrt{5}$\\\hline
			\end{tabular}
		\end{center}
	\end{table}
	
	In order to check if the set of almost quantum correlations can reproduce the maximally nonlocal quantum Hardy point, first we consider a line segment joining the PR box with the maximal quantum Hardy nonlocal point, say, $Q^{max}_{H}$, and then extend this line segment to the point $L_{H}$ on the local face. It turns out that the local point $L_{H}$ can be written as a convex combination of local deterministic points as follows,
	\begin{equation}
	L_H=\frac{9-\sqrt{5}}{38}\left(L_1+L_2+L_4+L_5\right)+\frac{1+2\sqrt{5}}{19}L_8.
	\end{equation} 
	Now we consider the set of all points on the line segment joining the PR box to the local point $L_H$, i.e., points generated by $\mu PR+ (1-\mu)L_H$, where $\mu\in [0,1]$. For these points, the success probability of Hardy's nonlocality argument is $p_H=\mu/2$. We know that the maximum success probability in quantum mechanics \cite{rabelo} is $(p_H)^{Q}_{*}=(5 \sqrt{5}-11)/2\simeq 0.09017$. We find the maximum value of $p_H$ in the set of almost quantum correlations $Q^{(1+ab)}$ on the considered line segment. For this we wrote a program in MATLAB, calling the function NPAHierarchy($\vec{p},1+ab$) from QETLAB~\cite{qetlab}, and computed the maximum value of $p_H$ for the set of almost quantum correlations. We find that  $(p_H)^{Q^{(1+ab)}}_{*}\simeq0.09024$. Our result shows a clear gap between the quantum and the almost quantum value.
	
	\section{Turning quantum voids into dimension witnesses}
	\label{sec:dim}
	So far we have focused only on the CHSH scenario. Indeed, the fact that any extremal correlation in the CHSH scenario can be obtained by projective measurements on qubit states \cite{mas1,mas2} was crucial in our proof of the quantum voids. That no longer holds true for other Bell scenarios, as, for example, in the generalization where each of the parties can measure one out of three observables, where we have a relevant class of Bell inequality \cite{I3322} given by
	\begin{eqnarray}
	& & \mathcal{I}_{3322}= p(0 0 \vert 0 0)+ p(00 \vert 01) + p(00 \vert 0 2) +
	p(00 \vert 10) \nonumber \\ 
	& & + p(0 0 \vert 1 1 )- p(00 \vert 12) +
	p(00 \vert 20) - p(00 \vert 2 1) \nonumber \\
	& & - p_B(0\vert 0) -  2 p_A(0\vert 0) - p_A(0\vert 1) \leq 0.
	\end{eqnarray}
	
	By imposing the condition $p_7=p(0, 0 \vert 1, 1) = 0$ and $p_4=p(0, 1 \vert 1, 0)=0$ 
	we know that this will correspond to a six-dimensional quantum void in the CHSH scenario. Also, it follows from the proof of the void that any qubit state respecting these constraints must be separable and then cannot violate any Bell inequality. Thus, any violation of the $\mathcal{I}_{3322}$ inequality, respecting the zero probability constraints, necessarily needs quantum states of dimension 3 or higher. That is precisely what we get by considering two-qutrit states and rank-1 projective measurements, that under the constraints $p(0 0 \vert 1 1) = 0$ and $p(0 1 \vert 1 0)=0$ achieve a violation of  $\mathcal{I}_{3322} \approx 0.2071$. 
	
	In other terms, the idea of a quantum void can be used to certify in a device-independent manner the minimum dimension of the physical system required to reproduce some given correlations. Clearly, in any experiment we will never see (with sufficiently many data points collected) that some of the probabilities are exactly equal to zero. However, we have strong numerical evidence for the robustness of the result. If the probabilities $p_7$ and $p_4$ are close to zero, the maximum violation of the CHSH inequality will also be small, and in contrast, violations of the $I_{3322}$ inequality can be significantly larger which leads to witnessing the dimension.

	\section{Discussion} \label{conclu}
	
	The understanding of the set of quantum correlations is important in both fundamental and practical applications. On one side, it allows us to witness the gap between classical, quantum, and postquantum predictions, thus giving insights on quantum theory itself. On the more practical level, it turns out that understanding the limits of physical theories is essential to come up with more efficient information protocols. In our case, the deeper one knows the boundary of quantum theory, the better one can explore quantum advantages and design enhanced quantum information protocols outperforming their classical counterparts. Here, we discuss a concept that analyzes and helps us to understand the set of Bell correlations, that of a quantum void, faces of a set of no-signaling correlations where all nonlocal correlations are postquantum in nature. Working in the simplest possible Bell scenario, we have given a full characterization of such quantum voids, their relations to known physical principles, and also pointed out a potential use for them as dimensional witnesses. In addition, among the faces which are not quantum voids, we studied the five-dimensional face derived from one set of Hardy's nonlocality conditions. We find that all physical principles (with our actual knowledge) fail to reproduce the quantum probability distribution, which gives the maximum success probability of Hardy's nonlocality argument in quantum mechanics.

	Very little was known about the faces of the no-signaling set and we hope our 
	results might motivate future research along this direction. We have focused here on the bipartite CHSH scenario and thus a natural extension would be to consider Bell scenarios with more parties, inputs, and outputs. At the same time, most of the research about the physical principles has so far focused on very limited regions of the NS set. This motivates us to pose the following question: Is there any principle capable of reproducing all quantum voids? As another interesting venue, we also highlight the use of quantum voids as  
	device-independent dimension witnesses.
	
	\begin{acknowledgments} 
		We thank Elie Wolfe, at the Perimeter Institute for Theoretical Physics,
		for his comments and suggestions on this work. We acknowledge support from the Brazilian ministries MCTIC, MEC, and the CNPq (PQ Grant No. $307172/2017-1$ and INCT-IQ), the Serrapilheira Institute (Grant No. Serra-1708-15763), and from John Templeton Foundation via Grant Q-CAUSAL No. 61084. (The opinions expressed in this publication are those of the authors and do not necessarily reflect the views of the John Templeton Foundation.) During the development of the manuscript, C.D. was also supported by a fellowship from the Grand Challenges Initiative at Chapman University.
	\end{acknowledgments}
	\appendix*

	\section{Characterization of all nonlocal faces of the $\mathbf{NL}$-simplex} \label{apdx}
	
	Here, we present a full characterization all the faces of the $\mathbf{NL}$ simplex into two categories: (i) those that are quantum voids, and (ii) those that are not quantum voids. For the cases which are quantum voids, the Fig.~\ref{fig3}, which follows from Proposition~$1$ of the main text, is sufficient to identify all the faces which are quantum voids, except for two of them for which we will provide a separate argument. The faces which are not quantum voids are demonstrated by showing the existence of quantum nonlocal points on these faces.     
	
	\subsection{Faces of dimension less than four}
	All faces of dimension less than four can be generated by assigning the value zero to at least five-free variable probabilities. In all such cases, one can check from the Fig.~\ref{fig3} that there will be always an edge connecting at least two free variable probabilities which are assigned the value zero. Therefore, all the considered types of faces are quantum voids. The number of such quantum voids of dimensions $0,~1,~2,~\mbox{and}~3$ are respectively $c^8_8=1,~ c^8_7=8,~c^8_6=28,~\mbox{and}~c^8_5=56$.

	\subsection{Four-dimensional faces} All four-dimensional faces are defined by assigning the value zero to any four free-variable probabilities. These are $c^8_4=70$ in numbers (each face being the convex hull of the PR box with four local vertices). 
	
	\emph{Case~1.} The four probabilities are chosen such that at least one probability is chosen from both $S1$ and $S2$, in this case, from Fig.~\ref{fig3} one can check that, there is always an edge between at least two among the four chosen probabilities. Thus this case leads to $68$ four-dimensional quantum voids.
	
	\emph{Case~2.} All four probabilities chosen to be zero are either from $S1$ or $S2$. Both the resulting faces also turn out to be quantum voids. One can prove this in two ways. First, by considering all possible pure qubit states and projective measurements \cite{mas1,mas2}, we maximized the Bell-CHSH expression (\ref{chsh}) under the given zero probability constraints to find that the maximum value is zero, i.e., the Bell-CHSH inequality is not violated, and hence there are no quantum nonlocal points on these faces. Second, we applied the known analytical condition \cite{npa1,npa2} for respecting the macroscopic locality principle, and then find that all the nonlocal points on these four-dimensional faces violate the macroscopic locality principle \cite{ml}, and hence there are no quantum nonlocal points on these faces.
	Our second proof uses Proposition~$2$ proved in the main text. For example, to prove that the four-dimensional face defined by assigning the value zero to all the four probabilities $\{p_1,p_2,p_7,p_8\}\in S_1$ (see Fig.~\ref{fig3}) is a quantum void, we consider only correlations on one line segment joining the PR box to the local point $L_c$ at the center of the local face, i.e., $\mu PR + \frac{1-\mu}{4}(L_3+L_4+L_5+L_6)$. We then showed that the macroscopic locality principle is violated for all $0<\mu\leq 1 $. Then, Proposition~$2$ in the main text implies that all nonlocal points on the considered four-dimensional face are postquantum points.
	
	Therefore, combining both the case studies, cases~$1$ and $2$, we can now conclude that all four-dimensional faces are quantum voids. 
	
	\subsection{Five-dimensional faces} All five-dimensional faces are defined by assigning the value zero to any three free-variable probabilities. These are $c^8_3=56$ in numbers (each face being the convex hull of the PR box with five local vertices). In contrast to all previous cases, here we find both types of five-dimensional faces: Some of these are quantum voids while other faces are not.
	
	\emph{Case~1.} All three probabilities chosen to be zero are either from $S1$ or $S2$. There are eight such cases and all the resulting five-dimensional faces contain quantum nonlocal points. We checked this by maximizing the Bell-CHSH expression [\cite{chsh}] under given probability constraints over all pure qubit states and projective measurements, and find that these maximum values are greater than the local (classical) bound. This clearly shows that there are many quantum nonlocal points on these five-dimensional faces.
	
	\emph{Case~2.} One probability is chosen from $S1$ and two from $S2$. There are $24$ such cases: Twenty cases lead to quantum voids (following from Fig.~\ref{fig3}), and the four remaining cases consist of quantum nonlocal points [examples follow when considering all types of Hardy nonlocality arguments given by conditions (\ref{h01})-(\ref{h04}) in Sec.~\ref{qHpoints} of the main text; all these Hardy nonlocality arguments have quantum solutions].
	
	\emph{Case~3.}  One probability is chosen from $S2$ and two from $S1$. This is similar to case~$2$: There are $24$ possibilities, out of which $20$ lead to quantum voids (follows from Fig.~\ref{fig3}), and four remaining cases consist of quantum nonlocal points [examples follow when considering all types of Hardy nonlocality arguments given by conditions (\ref{h01})-(\ref{h04}) in the Sec.~\ref{qHpoints} of the main text; all these Hardy nonlocality arguments have quantum solutions].
	
	To sum up, among all the five-dimensional faces, $40$ faces are quantum voids whereas $16$ faces contain quantum nonlocal points.

	\subsection{Six-dimensional faces} All six-dimensional faces are defined by assigning the value zero to any two free-variable probabilities. These are $c^8_2=28$ in numbers (each face being the convex hull of the PR box with six local vertices). Here, too, we find both types of six-dimensional faces: Some of these are quantum voids whereas others are not.
	From Fig.~\ref{fig3} one can see that eight cases lead to quantum voids. In all the remaining $20$ cases, we find quantum nonlocal points, and to show this we simply note that from all the examples of five-dimensional faces containing quantum nonlocal points, one can provide examples of quantum nonlocal points on all these six-dimensional faces.  
	
	\subsection{Seven-dimensional faces}  All seven-dimensional faces are defined by assigning the value zero to any one free-variable probability. These are $c^8_1=8$ in numbers (each face being the convex hull of the PR box with seven local vertices). None of these eight faces is a quantum void [examples follow when considering all types of Hardy nonlocality argument given by conditions (\ref{h01})-(\ref{h04}) in the Sec.~\ref{qHpoints} of the main text; all these Hardy nonlocality arguments have quantum solutions).
	
	We summarize the number of nonlocal faces of the $\mathbf{NL}$ simplex that are quantum voids in Table~\ref{table2}.
	
	\begin{table}[h!]
		\begin{center}
			\caption{Number of nonlocal faces which are quantum voids.}
			
			\label{table2}
			\begin{ruledtabular}
				\begin{tabular}{ccc}
					\textbf{Dimension} &\#~\textbf{Faces} & \#~\textbf{Q-voids}\\
					\colrule
					\hline
					0 & 1~~~ & 1~~~~~~\\
					1 & 8~~~ & 8~~~~~~\\
					2 & 28~~~ & 28~~~~~~\\
					3 & 56~~~ & 56~~~~~~\\
					4 & 70~~~ & 70~~~~~~\\
					5 & 56~~~ & 40~~~~~~\\
					6 & 28~~~ & 8~~~~~~\\
					7 & 8 ~~~& 0~~~~~~\\
				\end{tabular}
			\end{ruledtabular}
		\end{center}
	\end{table}


\begin{thebibliography}{99}
\bibitem{lhv} There is no loss of generality in considering a deterministic local-hidden-variable model, because for a given correlation, any probabilistic local-hidden-variable model $p(a,b\vert x,y)=\sum_{\lambda} p(\lambda) p^{A}(a\vert x,\lambda)p^{B}(b\vert y,\lambda)$ can be always transformed to a deterministic local-hidden-variable model $p(a,b\vert x,y)=\sum_{\tilde{\lambda}} \tilde{p}(\tilde{\lambda}) D^{A}(a\vert x,\tilde{\lambda})D^{B}(b\vert y,\tilde{\lambda})$, where $D^{*}(*\vert *,*)$ are deterministic functions taking a definite value, either $0$ or $1$, for any given measurement and local hidden-variable.
\bibitem{hypograph} The hypograph of a function $f: D\rightarrow \mathbb{R}$, where the domain $D\subseteq \mathbb{R}^n$, is the set of points lying on or below its graph, i.e., $\mbox{hyp}f=\{(x,\mu): x\in D, \mu\in \mathbb{R}, \mu \leq f(x) \}\subseteq \mathbb{R}^{n+1}$ (see p. 9 of Ref.~\cite{book}). A crucial property is that a function $f$ is concave if and only if its hypograph is a convex set (see p. 187 of Ref.~\cite{book}).
\end{thebibliography}
\end{document}